\documentclass[12pt]{article}
\usepackage{times}
\usepackage{geometry}
\usepackage[utf8]{inputenc}
\usepackage{enumitem}
\usepackage{ragged2e}
\newlist{thematic}{itemize}{8}
\setlist[thematic]{label=$\square$}
\usepackage{pifont}

\usepackage{graphics,graphicx}

\usepackage{float}
\usepackage[colorlinks = true,
            linkcolor = blue,
            urlcolor  = blue,
            citecolor = blue,
            anchorcolor = blue]{hyperref}
\usepackage{epsfig}
\usepackage{lineno}
\usepackage{natbib}
\usepackage{amsmath}
\usepackage{amssymb}
\usepackage[normalem]{ulem}
\usepackage{multicol}

\newcommand{\fermi}{{\it Fermi}}
\newcommand{\swift}{{\it Swift}}

\newcommand{\nustar}{\textit{NuSTAR}}

\newcommand{\msun}{$M_{\odot}$}

\newcommand{\gm}{$\gamma$}

  \renewenvironment{thebibliography}[1]{%
    \begin{oldthebibliography}{#1}%
      \setlength{\parskip}{0ex}%
      \setlength{\itemsep}{0ex}%
  }%
  {%
    \end{oldthebibliography}%
  }

\begin{document}
\pagestyle{plain}
\pagenumbering{arabic}

\raggedright
\huge
Astro2020 Science White Paper \linebreak
 
Supermassive black holes at high redshifts \linebreak
 \normalsize

\noindent \textbf{Thematic Areas:} \hspace*{60pt} \linebreak
$\text{\rlap{$\checkmark$}}\square$ Formation and Evolution of Compact Objects \hspace*{31pt} $\text{\rlap{$\checkmark$}}\square$ Cosmology and Fundamental Physics \linebreak
  
\textbf{Principal Authors:}

Name:	Vaidehi S. Paliya
 \linebreak						
Institution:  Clemson University \& Deutsches Elektronen Synchrotron DESY, Germany
 \linebreak
Email: vaidehi.s.paliya@gmail.com
 \linebreak
Phone:  +49 175 2112107
 \linebreak
 \\
 Name:	Marco Ajello
 \linebreak						
Institution:  Clemson University
 \linebreak
Email: majello@g.clemson.edu
 \linebreak
Phone:  +1 650 8049042
 \linebreak
 \\
 Name:	Lea Marcotulli
 \linebreak						
Institution:  Clemson University
 \linebreak
Email: lmarcot@g.clemson.edu
 \linebreak
Phone:  +1 864 6507068
 \linebreak

 \textbf{Co-authors:} J.~Tomsick, Space Sciences Lab, University of California Berkeley; J.~S.~Perkins, NASA/GSFC; E.~Prandini, University of Padova, Italy; F.~D'Ammando, INAF-IRA Bologna; A.~De Angelis,	INFN and INAF Padova, Italy; D.~Thompson, NASA/GSFC; H.~Li, LANL; A.~Dom{\'i}nguez, GAE and IPARCOS, Universidad Complutense de Madrid; V.~Beckmann, CNRS / IN2P3; S.~Guiriec, GWU/NASA GSFC; Z.~Wadiasingh, NASA/GSFC; P.~Coppi, Yale University; J.~Patrick~Harding,	LANL; M.~Petropoulou, Princeton University; J.~W.~Hewitt, University of North Florida; R.~Ojha, UMBC/NASA GSFC; A.~Marcowith, Laboratoire Univers et Particules de Montpellier; M.~Doro, University and INFN Padova; D.~Castro, Center for Astrophysics, Harvard \& Smithsonian; M.~Baring,	Rice University; E.~Hays, NASA GSFC; E.~Orlando, Stanford University; S.~Guiriec, GWU/NASA GSFC; V.~Bozhilov, Sofia University; J.~Tomsick	UC Berkeley; I.~Agudo,	IAA-CSIC (Spain); T.~Venters, NASA GSFC; J.~McEnery, NASA/GSFC; L.~The, Clemson University; D.~Hartmann, Clemson University; S.~Buson, Universit\"{a}t W\"{u}rzburg, University of Maryland, Baltimore County, USA; Francesco Longo, University of Trieste and INFN, Trieste; Dario Gasparrini, SSDC and INFN, Roma Tor Vergata.\linebreak

\textbf{Abstract: MeV blazars are the most luminous persistent sources in the Universe and emit most of their energy in the MeV  band.
These objects  display very large jet powers and accretion luminosities and are known to host black holes with a mass often 
exceeding 10$^9$\msun. 
An MeV survey, performed by a new generation MeV telescope which will bridge the entire energy and sensitivity gap between the current generation of hard X-ray and \gm-ray instruments, will detect $>$1000 MeV blazars up to a redshift of $z=5-6$. Here we show that this would allow us: 1) to probe the formation and growth mechanisms of supermassive black holes at high redshifts, 2) to pinpoint the location of the emission region in powerful blazars, 3) to determine how accretion and black hole spin interplay to power the jet.}
\linebreak

\vspace{-0.1cm}
\underline{Projects/Programs Emphasized:} \\
\begin{enumerate}
\item All-sky Medium Energy Gamma-ray Observatory (AMEGO)\\
\url{https://asd.gsfc.nasa.gov/amego/}\\
\vspace{-0.2cm}
\item Lunar Occultation Explorer (LOX)  \\
\url{https://www.hou.usra.edu/meetings/deepspace2018/pdf/3094.pdf}\\
\vspace{-0.2cm}
\item Compton Spectrometer and Imager (COSI) \\
\url{http://cosi.ssl.berkeley.edu}\\
\vspace{-0.2cm}
\item e-ASTROGAM \\
\url{http://eastrogam.iaps.inaf.it}\\
\vspace{-0.2cm}
\item adEPT \\
\url{http://ui.adsabs.harvard.edu/#abs/2014APh....59...18H}\\
\end{enumerate}

\newpage
\justify

\section{Introduction}
Black holes as massive at $10^{9-10}$\,\msun\ have been discovered all the way up to redshift $z=6-7$ (Wu et al. 2015, Mortlock et al. 2011, Ba{\~n}ados et al. 2018). Their existence poses crucial questions on our understanding of black hole formation and growth. Indeed, already at $z>4$, the Universe's age is barely compatible with the time it takes to grow a 10$^{9-10}$ \msun~black hole just by accretion (at the Eddington limit) starting from a stellar mass black hole. 
Thus, the detection of a large number of massive black holes early on  implies that: 1) they could have grown from a massive (10$^{5-6}$\msun) seed (formed through, e.g., the collapse of a massive molecular torus following the merging of the giant protogalaxies; Mayer et al. 2010),  2) experienced sustained super-Eddington accretion (Volonteri \& Rees 2005), or 3) have experienced a large number of merging events.

\begin{figure*}[h]
\begin{center}
\begin{tabular}{lll}
\hspace{-0.5cm}
\includegraphics[scale=0.27]{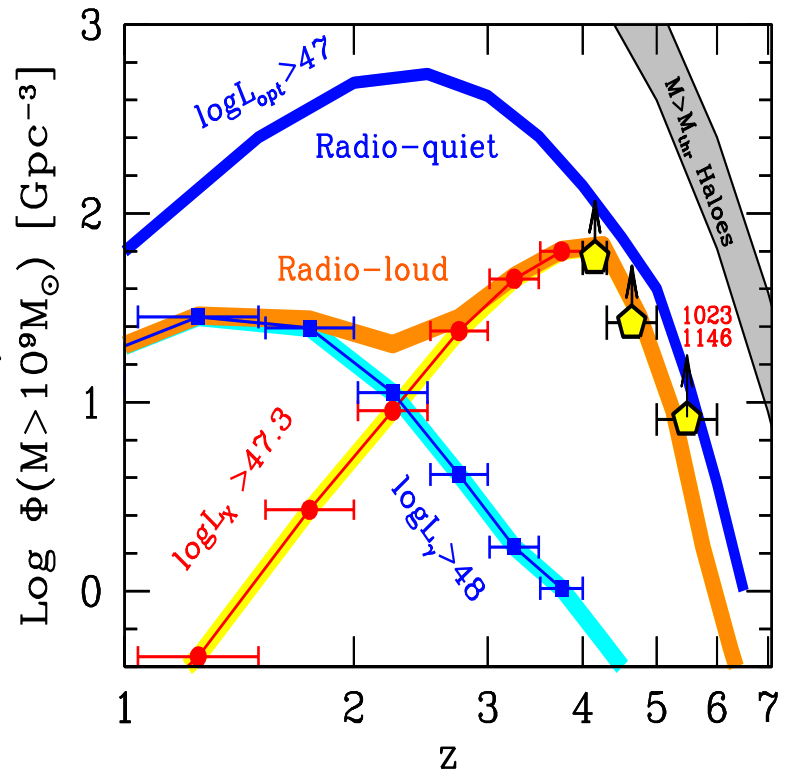} 
\hspace{+0.5cm}
\includegraphics[scale=0.33]{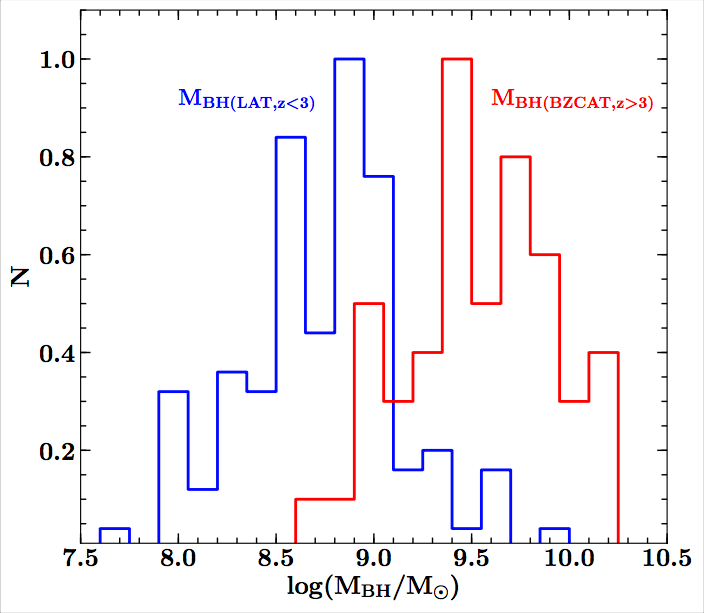}
\end{tabular}
\end{center}
\vspace{-0.4cm}
\caption{\footnotesize 
{\footnotesize{\bf Left:} Space density of billion-solar-mass black holes hosted in radio-quiet (blue line) and radio-loud quasars (orange line). The cyan and yellow lines represent number densities derived from the \fermi-LAT (Ajello et al. 2012) and \swift-BAT detected blazars (Ajello et al. 2009), respectively. Adopted from Sbarrato et al. (2015). {\bf Right:} Black hole mass distributions for blazars detected in the GeV band (blue, typically at $z<3$) and the MeV blazars (red).}
\label{fig:1}}
\end{figure*}

Quasars are the most luminous active galactic nuclei (AGN) known, and, as such, they enable unparalleled studies of the Universe at the earliest cosmic epochs. The space density of massive black holes ($M>10^9$~\msun) hosted in radio-quiet quasars\footnote{Formally, a source is called radio-quiet if $R<10$, where radio-loudness parameter $R$ is the ratio of the rest-frame 5 GHz to optical B-band flux density (Kellermann et al. 1989). A radio-loud object has $R>10$.} peaks at $z\approx2$ (Hopkins et al. 2007, see also blue line in the right panel of Fig.~\ref{fig:1}). However, massive black holes in radio-loud quasars (objects with strong jetted radio emission) are present at much earlier epochs, close to $z=4-5$ (orange line in the right panel of Fig.~\ref{fig:1}, Sbarrato et al. 2015).
{\bf This points to the fact that relativistic jets may play a crucial role in rapid black hole growth at these early stages of the Universe.}
The observation of  high-redshift jetted quasars will thus allow us to probe the formation and growth mechanisms of massive black holes at the dawn of the Universe.

Blazars, radio-loud AGN having relativistic jets oriented close to the line of sight, have been detected  to $z>5$ (Romani et al. 2004). Already at moderately high redshifts ($z>3$), blazars host massive ($>$10$^9$ \msun) black holes and display extremely luminous ($>$10$^{46}$ erg s$^{-1}$) accretion disks (e.g., Ghisellini et al. 2010). Most remarkably, these powerful objects emit most of their energy output in the MeV band and have thus been dubbed `MeV' blazars. These sources, among all other blazars, tend to host the most massive black holes and are also found at the highest redshifts, as seen in Fig.~\ref{fig:1} (right panel). Therefore a survey of `MeV' blazars, as explained below, will allows us to constrain the space density of massive black holes in the early Universe.

\section{Importance of the MeV Band}
\vspace{-0.2cm}

The most powerful and distant blazars emit the majority of their bolometric output between 0.2\,MeV and 10\,MeV, making the MeV band the optimal one to search for them. 
So far, the detection of such elusive monsters has been achieved using \gm-ray (\fermi-Large Area Telescope, LAT, $\rm[20\,MeV,2\,TeV]$, Atwood et al. 2009) or hard X-ray (\swift-Burst Area Telescope, BAT, $\rm[14,195\,keV]$, Gehrels et al. 2004; Nuclear Spectroscopic Telescope Array, \nustar, $\rm[3,80\,keV]$, Harrison et al. 2013) observatories, or a combination of the two. Peaking at MeV energies, these sources become harder to detect in the \gm-ray band, due to both the softening of the spectrum ($\Gamma_{\gamma}>2.7$) as well as instrument sensitivity threshold. Only 8 blazars have been detected by the LAT  beyond redshift 3 (Marcotulli et al. 2017, Ackermann et al., 2017), the farthest being at $z=4.31$. On the other hand, the shallow sensitivity of the BAT has allowed us to find only a handful of the brightest sources (Oh et al., 2018). Even if the exceptional sensitivity of \nustar~allows for the detection of fainter sources, its pointing strategy does not permit a population study. Moreover, hard X-ray observations alone of these sources (without \gm-ray ones) are not able to pinpoint the peak of the emission. This, in turn, is of fundamental importance to measure jet parameters, such as the total jet power and the bulk Lorentz factor ($\Gamma$). Moreover, since their jetted radiation is Doppler boosted and collimated within an angle $\sim 1/\Gamma$, {\bf for each detected blazar there are 2$\Gamma^2$ ($\sim200-450$, for typical $\Gamma\sim10-15$) radio-loud quasars with similar intrinsic properties and  jets  pointed elsewhere.} Thus, the detection of even a few blazars and an accurate measurement of $\Gamma$ can constrain in a sensitive way the space density of massive black holes at high redshifts (see Fig.~\ref{fig:1} and Sbarrato et al. 2015, Ackermann et al. 2017).
An all-sky MeV instrument is therefore the necessary bridge in both sensitivity and energy coverage between the existing state-of-the-art instruments in the other high-energy bands. Directly sampling blazars' peak emission, it is the best solution to uncover the population of extreme high-redshift blazars, to precisely constrain their jet properties and to uncover the most massive black holes in the Universe and their connection to jets.



\section{Science with MeV Observations of Powerful blazars}

\begin{figure*}[t]
\begin{center}
\begin{tabular}{ll}
\hspace{-1.0cm}
\includegraphics[scale=0.44]{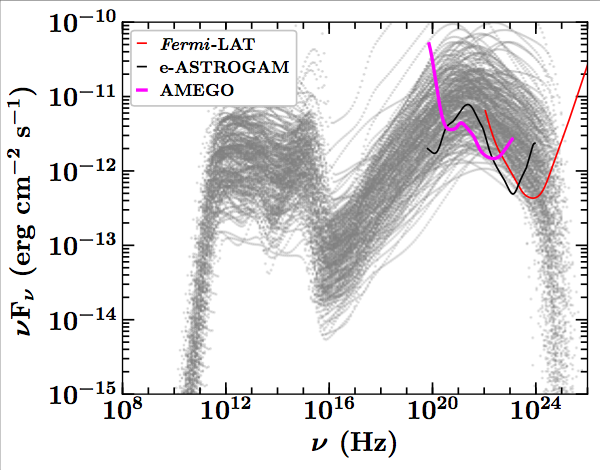}\
\hspace{-0.5cm}
\includegraphics[scale=0.4]{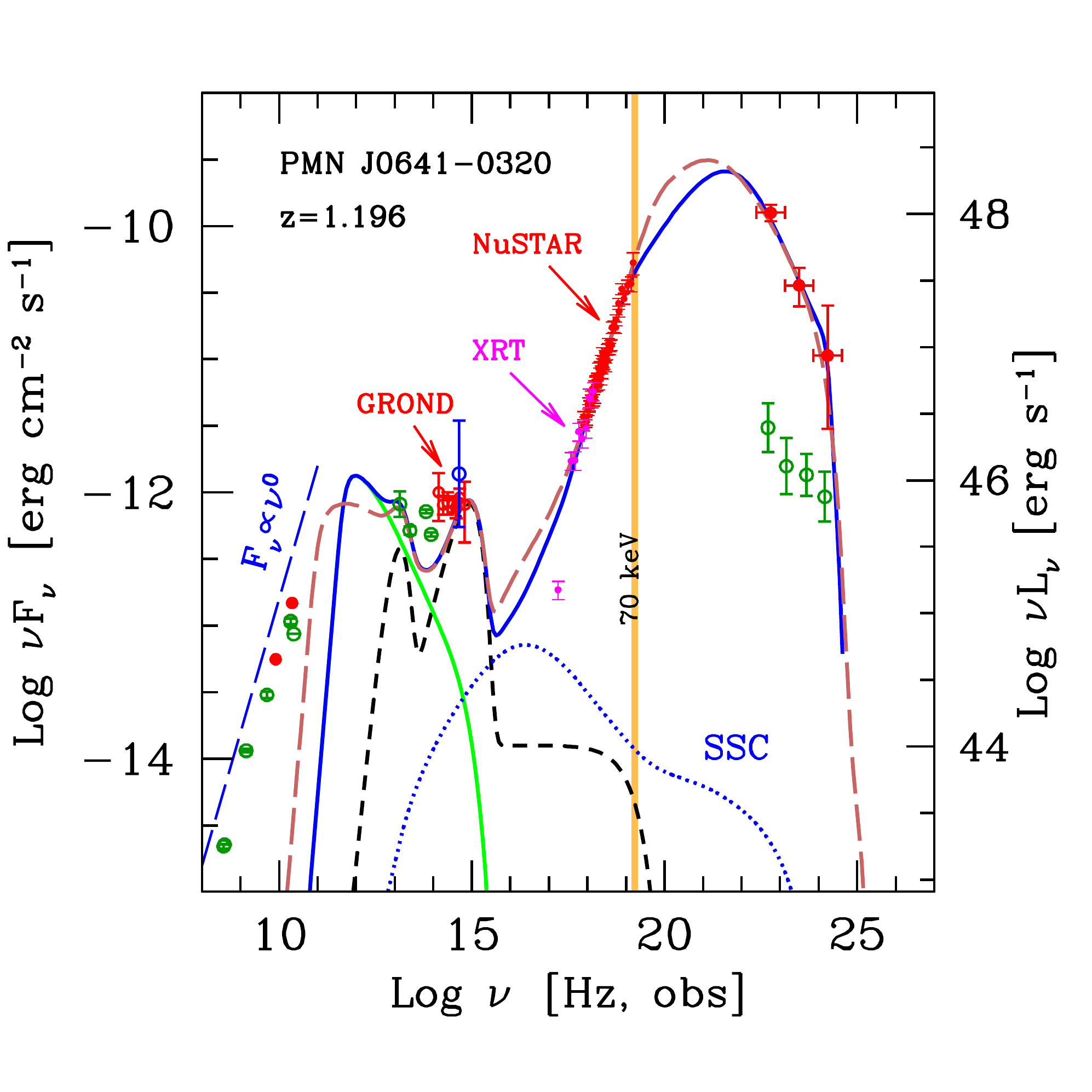}\ 
\end{tabular}
\end{center}
\vspace{-0.9cm}
\caption{\footnotesize {\bf Left:} The broadband SEDs of \gm-ray detected blazars studied in Paliya et al. (2017). Overplotted are the \fermi-LAT sensitivity in 10 years of observation (red) and the 3$\sigma$ sensitivity plot for one year of exposure of the
proposed all-sky MeV missions AMEGO (pink) and e-ASTROGAM (black).
{\bf Right:} Modeled SED of a MeV blazar PMN J0641$-$0320 ($z=1.196$). Blue solid and brown dashed lines represent the modeling done by assuming the seed photons for inverse Compton scattering (so-called external Compton or EC process) originated from the BLR or dusty torus, respectively. Note the bright hard X-ray emission observed from \nustar~which, along with \fermi-LAT spectra, has accurately constrained the inverse Compton peak. Adopted from Ajello et al. (2016).} 
\label{fig:3}
\end{figure*}

An all-sky MeV mission like AMEGO, LOX, e-ASTROGAM or adEPT, with a continuum sensitivity in the 0.1-10\,MeV energy band of 10$^{-11}$\,erg cm$^{-2}$ s$^{-1}$, will detect $\approx$2000 blazars. Half of these objects will be detected at redshift $>2$ and, importantly, $\sim$50 will be found between redshift of 5 and 6 (Ajello et al. 2009). 
Moreover, LAT-known blazars will be uncovered by such instruments. As can be seen from Fig.~\ref{fig:3}, if these sources display steeply falling spectral energy distribution (SED) in the GeV band, they are much brighter in the MeV band and will be easily captured by these MeV missions.
\\\\
{\bf Evolution and Massive Black Hole Growth:} This sample will be used to study the evolution of the most powerful blazars, which are known to experience extremely strong positive evolution (i.e. space density growing quickly with increasing redshift) and to trace the evolution of the most massive elliptical galaxies (Ajello et al. 2009).

As shown in the right panel of Fig.~\ref{fig:1}, we expect most of the blazars at $z\geq$3 to host supermassive black holes with a mass in excess of 10$^9$\msun. Therefore, this set of blazars can be used to measure the space density of billion solar mass black holes at high redshifts. 
Already at $z=4$ this function for radio-quiet and radio-loud systems is the same, 
implying that the radio-loud phase (likely triggered by a major merger) may be essential for quick black hole growth. This survey of powerful blazars will allow us to extend this measurement beyond redshift of $z=4$, where it becomes virtually impossible to grow a black hole purely by accretion at the Eddington limit.
\\\\
{\bf Location and Size of the Emission Region:}
The high-energy emission of FSRQs is typically interpreted as owing to inverse Compton scattering on either the photons of the broad line region (BLR) or those of the torus (Begelman \& Sikora 1987, Paliya et al. 2017). Since the characteristic energy of BLR photons ($\sim$10 eV) is larger compared to the torus ones ($\sim$0.1 eV), inverse Compton scattered emission will peak at higher frequencies if upscattering BLR photons rather than photons from the torus.
The peak frequency is particularly sensitive to (and scales linearly with)  the energy of the external photon fields and can thus be used to discriminate between different external field components and whether the emission region resides within the BLR or beyond (see Fig.~\ref{fig:3}, right). Virtually all the high-energy peaks of known FSRQs fall in the MeV regime and, as of now, sensitive observations at these wavelengths are crucially lacking. 
Moreover, in a conical jet scenario, an emission region located within the BLR is $\sim$5$-$10 times smaller than that one located outside of it and within the torus. Thus, the typical timescale for the variability of FSRQs in the MeV band will strongly be affected by the location, hence size, of the emission region (with way faster variability in the BLR scenario with respect to the torus one).
An MeV mission will thus yield two critical observables to determine the location and size of the emission region in FSRQs.
\\\\
{\bf Accretion-Jet Connection:} It has been long postulated that jets are powered by the accretion disk (Rawlings \& Saunders, 1991). In recent years, it has come to attention that, for jetted quasars at $z<3$, an extra source of power (e.g. black hole spin) is also needed to energize these jets (Ghisellini et al., 2014). Whether this holds true beyond $z=3$ is still an open issue. However, these theories strongly rely on accurate measurements of both the accretion disk power as well as the jet one. Though the former can be constrained from optical spectroscopy (e.g., Shaw et al., 2012), it is tedious to precisely measure the latter due to lack of sensitive hard X-ray-to-\gm-ray (i.e., MeV band) observations. To this goal, the MeV regime is crucial. Firstly because these powerful blazars radiate a major fraction of their bolometric luminosity in this energy range. Secondly, measurements in this band are needed in order to accurately parametrize the underlying particle population. The synchrotron peak in MeV blazars is typically located at radio to sub-mm wavelengths. Because of synchrotron self-absorption, it is impossible to characterize the spectral shape of the underlying electron energy distribution (EED) from the synchrotron emission alone. In such MeV blazars, the EED can be constrained using hard X-ray and $\gamma$-ray data, which (in EC scenario\footnote{The large Compton dominance in powerful MeV blazars indicates a relatively low magnetic field and thus a negligible contribution from synchrotron self Compton at X-rays.}) constrain the low-energy and high-energy electron population, respectively. In particular, the spectral index before the break energy of the EED\footnote{The EED is typically modeled as broken power law.} (as measured from MeV observations, Fig.~\ref{fig:3}, right panel) and the minimum energy of the EED control the total amount of radiating electrons, thus also the number of cold protons, assuming one cold proton per electron (Celotti \& Ghisellini, 2008) which dominate the total power of the jet. Therefore, the knowledge of the hard X-ray-to-MeV energy band is crucial to
accurately determine the total jet power. A comparison of the accretion disk luminosity with the jet power will enable us to determine whether jets are mainly powered by accretion or there are also other reservoir of energy, such as black hole spin. Most importantly, we will be able to study this connection for a whole population of high-redshift sources, possibly unveiling new jet physics.

\vspace{0.5cm}
\newpage
\noindent {\large {\bf References}:}
 \begin{multicols}{2}
{\noindent Atwood, et al. 2009, ApJ, 697, 1071 \\
Ackermann et al., 2017, ApJL, 837, 5 \\
Ajello et al., 2009, ApJ, 699, 603 \\
Ajello et al., 2012, ApJ, 751, 108  \\
Ba{\~n}ados et al., 2018, Nature, 553, 473 \\
Begelman \& Sikora, 1987, ApJ, 322, 650 \\
Blandford \& Rees, 1978, Phys. Scr, 17, 265 \\
Bloemen et al., 1995, A\&A, 293, 1 \\
Celotti \& Ghisellini, 2008, MNRAS, 385, 283 \\
Dermer, 1995, ApJ, 446, 63 \\
Dermer \& Schlickeiser, 1993, ApJ, 416, 458 \\
Gehrels, et al. 2004, ApJ, 611, 1005 \\
Ghisellini et al., 1998, MNRAS, 301, 451 \\
Ghisellini et al., 2010, MNRAS, 405, 387 \\
Ghisellini et al., 2014, Nature, 515, 376 \\
Harrison et al. 2013, ApJ, 770, 103 \\
Hopkins et al., 2007, ApJ, 654, 731 \\
Kellermann et al., 1989, AJ, 98, 1195 \\
Mayer et al., 2010, Nature, 466, 1082 \\
Marcotulli et al. 2017, ApJ, 839, 96\\
Mortlock et al., 2011, Nature, 474, 616\\
Paliya et al., 2017, ApJ, 851, 33 \\
Rawlings \& Saunders, 1991, Nature, 349, 138 \\
Romani et al., 2004,  ApJL, 610, 9 \\
Sbarrato T., et al. 2015, MNRAS, 446, 2483 \\
Shaw et al., 2012, ApJ, 748, 49 \\
Volonteri \& Rees, 2005, ApJ, 633, 624 \\
Volonteri M., et al., 2011, MNRAS, 416, 216 \\
Wu et al., 2015, Nature, 518, 512 \\
}
\end{multicols}

\end{document}